\documentclass[
 reprint,
 amsmath,amssymb,
 apl
]{revtex4-1}

\usepackage{graphicx}
\usepackage{dcolumn}
\usepackage{bm}
\usepackage{siunitx}
\usepackage[hidelinks]{hyperref}
\usepackage[utf8]{inputenc}
\newcommand{\iu}{{i\mkern1mu}}

\def\figureautorefname{Fig.}
\def\equationautorefname{Eq.}
\newcommand{\Autoref}[1]{%
  \begingroup%
  \def\chapterautorefname{Chapter}%
  \def\sectionautorefname{Section}%
  \def\subsectionautorefname{Sub--Section}%
  \def\figureautorefname{Figure}%
  \def\tableautorefname{Table}%
  \def\equationautorefname~#1{Equation~(#1)}%
  \autoref{#1}%
  \endgroup%
}

\begin{document}
\preprint{APS/123-QED}

\title{
Fast and accurate functionalization of opaque conductive samples with single nano particles
}
\author{Niko Nikolay$^{1,2}$}
\author{Nikola Sadzak$^{1,2}$}
\author{Alexander Dohms$^{1,2}$}
\author{Boaz Lubotzky$^3$}
\author{Hamza Abudayyeh$^3$}
\author{Ronen Rapaport$^3$}
\author{Oliver Benson$^{1,2}$}
\affiliation{
 $^1$ AG Nanooptik, Humboldt Universität zu Berlin,
 Newtonstraße 15, D-12489 Berlin, Germany
}
\affiliation{
 $^2$ IRIS Adlershof, Humboldt Universität zu Berlin,
 Zum Großen Windkanal 6, 12489 Berlin, Germany
}
\affiliation{
 $^3$ The Racah Institute of Physics, The Hebrew University of Jerusalem,
 Jerusalem 9190401, Israel
}

\date{\today}
\begin{abstract}
Single quantum emitters coupled to different plasmonic and photonic structures are key elements for integrated quantum technologies. In order to fully exploit these elements, e.g. for quantum enhanced sensors or quantum repeaters, a reliable fabrication method as enabling technology is crucial. In this work, we present a method that allows for positioning of individual nano crystals containing single quantum light sources on non-transparent conductive samples with sub-micrometer precision. We induce long-range electrostatic forces between an atomic force microscope (AFM) tip, which carries a nano particle, and the target surface. This allows for mapping of the target area in non contact mode. Then, the placement site can be identified with high accuracy without any tip approach, eliminating the risk of a particle loss. We demonstrate the strength of the method by transferring a diamond nano crystal containing a single nitrogen-vacancy defect to the center of a micrometer-sized silver bullseye antenna with nanometer resolution. Our approach provides a simple and reliable assembling technology for positioning single nano objects on opaque substrates with high reproducibility and precision.
\end{abstract}

\maketitle

\begin{figure}[ht]
 \includegraphics[width=\columnwidth]{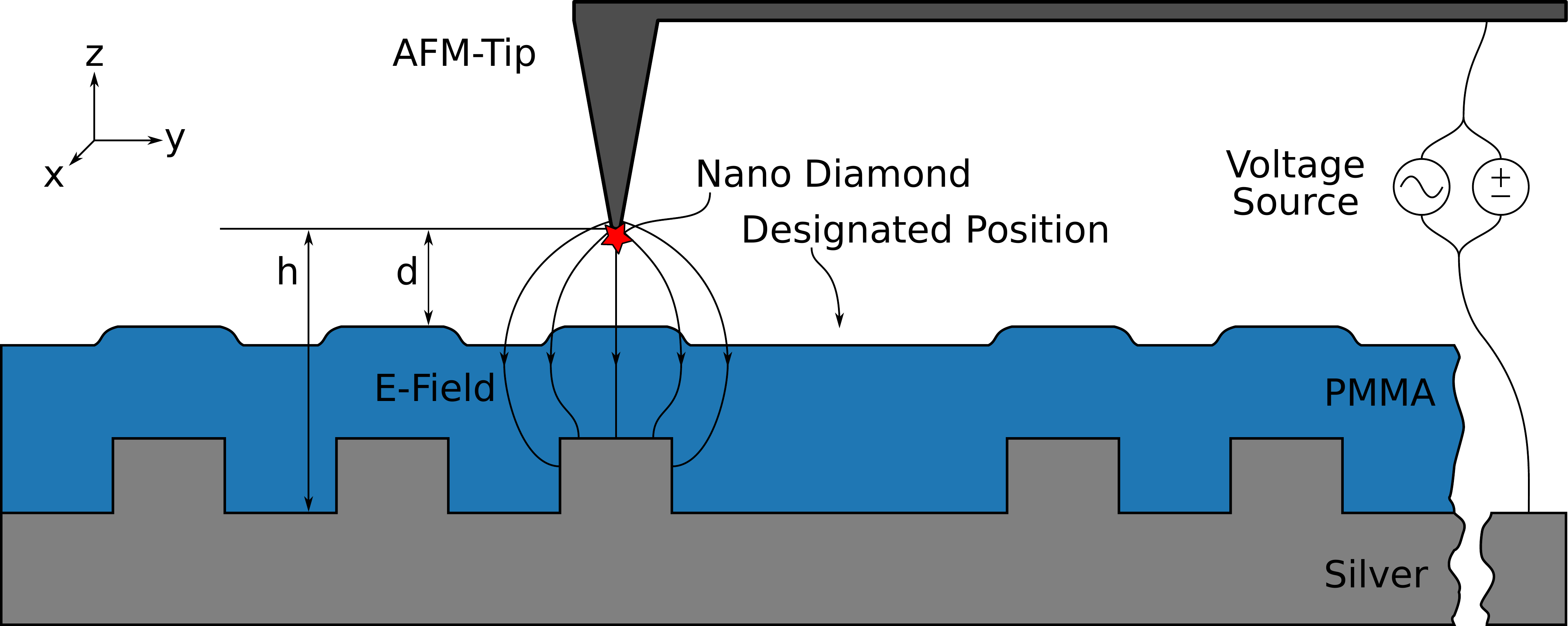}%
 \caption{\textit{Schematic representation of the functionalization approach combining the pick and place technique with electrostatic force imaging.} As exemplary structure we chose a nano diamond to be placed on a bullseye antenna. A cross section through the cylindrical antenna made of silver and PMMA is shown. While the AFM cantilever holds the nano diamond, an applied voltage creates electric fields as indicated. The electrostatic force provides a feedback and reveals the topography of the substrate when scanning the cantilever. Excellent resolution is obtained even with large (up to 2 micrometer) tip-substrate distances. In this way the risk of loosing the nano diamond while approaching a designated location on the sample is highly diminished.}
 \label{fig:schematics}
\end{figure}

An ideal platform to study light-matter interaction at the fundamental level consists of single quantum emitters coupled to photonic and plasmonic elements \cite{Akimov2007,Wolters2010,Benson2011a,Andersen2018}. Such elements are also needed to realize quantum interfaces between stationary and flying qubits operating with near unity efficiency \cite{Bernien2013a}. Reaching the required nanometer precision for optimum coupling is still a challenge. Approaches for different scenarios have been developed. Ion implantation in diamond nano structures was successful to create optically active defect centers at well defined locations \cite{Meijer2005,Meijer2008,Sipahigil2016a}. In a complementary approach, first single emitters as quantum dots or defects in diamond nano crystals (nano diamonds) are identified and subsequently a structure is built around them \cite{Gschrey2013, Shi2016}. In other cases emitters are spin-coated on masks exposing just the part of the sample that needs to be functionalized \cite{Bermudez-Urena2015}; or the entire surface of a sample is covered with emitters and misplaced ones are etched away \cite{Harats2017}. Another promising technique is the pick and place method. Here, single nano particles containing quantum emitters are transferred from substrate to substrate by employing an atomic force microscope (AFM) \cite{Schell2011} combined with a confocal scanning microscope (CSM). The method is highly accurate and deterministic and it also allows for pre-characterization of the luminescent particles. Moreover, the placement is not final, several iterations can be performed by nano manipulation if required. Unlike most other methods, neither a vacuum environment nor large and expensive electron beam lithography systems are required. However, the positioning precision relies on the fluorescence feedback observed via a CSM accessing the surfaces from below the substrate. Many important substrates for integrated quantum photonics, however, are opaque: dielectric waveguides and resonators are fabricated on silicon, plasmonic structures are based on metal films. In this case the pick and place method may still be applied \cite{Wolters2010,Andersen2018}, but it is generally not possible to place the fluorescent nano crystal with sub-micrometer precision in a single shot. Here we present a fast and versatile technique allowing for high accuracy placement even on opaque structures. In order to demonstrate our method we chose a highly relevant architecture, i.e. a nano diamond containing a single nitrogen vacancy defect center (NV) coupled to a light collecting optical antenna. Such an element renders particularly useful as a quantum light source \cite{Shi2016}, or for a fast read-out of the NV’s electron spin and thus faster and more efficient magnet field sensing \cite{Schirhagl2014}. We use a Polymethylmethacrylate (PMMA) coated bullseye antenna \cite{Harats2017} whose cross section is schematically shown in \autoref{fig:schematics} together with the conceptual idea of our approach. The antenna consists of a slab waveguide and concentrically arranged circular metal rings. The emitter couples to the slab waveguide that distributes the radiation over the metal grating. Each metal ring coherently scatters the emission, resulting in interference and a collimated light beam. A detailed discussion of the antenna’s working principle can be found in Ref. \cite{Abudayyeh2017}.\\
The functionalization approach starts with picking up a pre-characterized nano diamond of some ten nanometers in size hosting a single NV using a commercial AFM similar to what is described in Ref. \cite{Schell2011}. In order to transfer the nano particle to a designated position on a structure on top of an opaque substrate with sub-micrometer precision requires knowledge of the structure's topography. In principle this could be gained by scanning the AFM-tip across the substrate. However, with a functionalized tip carrying the nano particle, which is only attached by surface adhesion, this is impossible in regular AFM scanning modes because the particle would be detached too easily during scanning. Even the approach of the tip to the surface may lead to a loss of the nano diamond. To overcome this challenge, a non-contact approach and imaging principle must be used. In our method, we induce long range Coulomb interactions by applying a voltage between conductive tip and sample. This can be seen as a modified Kelvin Probe measurement, where the tip-surface distance is not kept constant \cite{Melitz2011}.

\begin{figure}[ht]
 \includegraphics[width=\columnwidth]{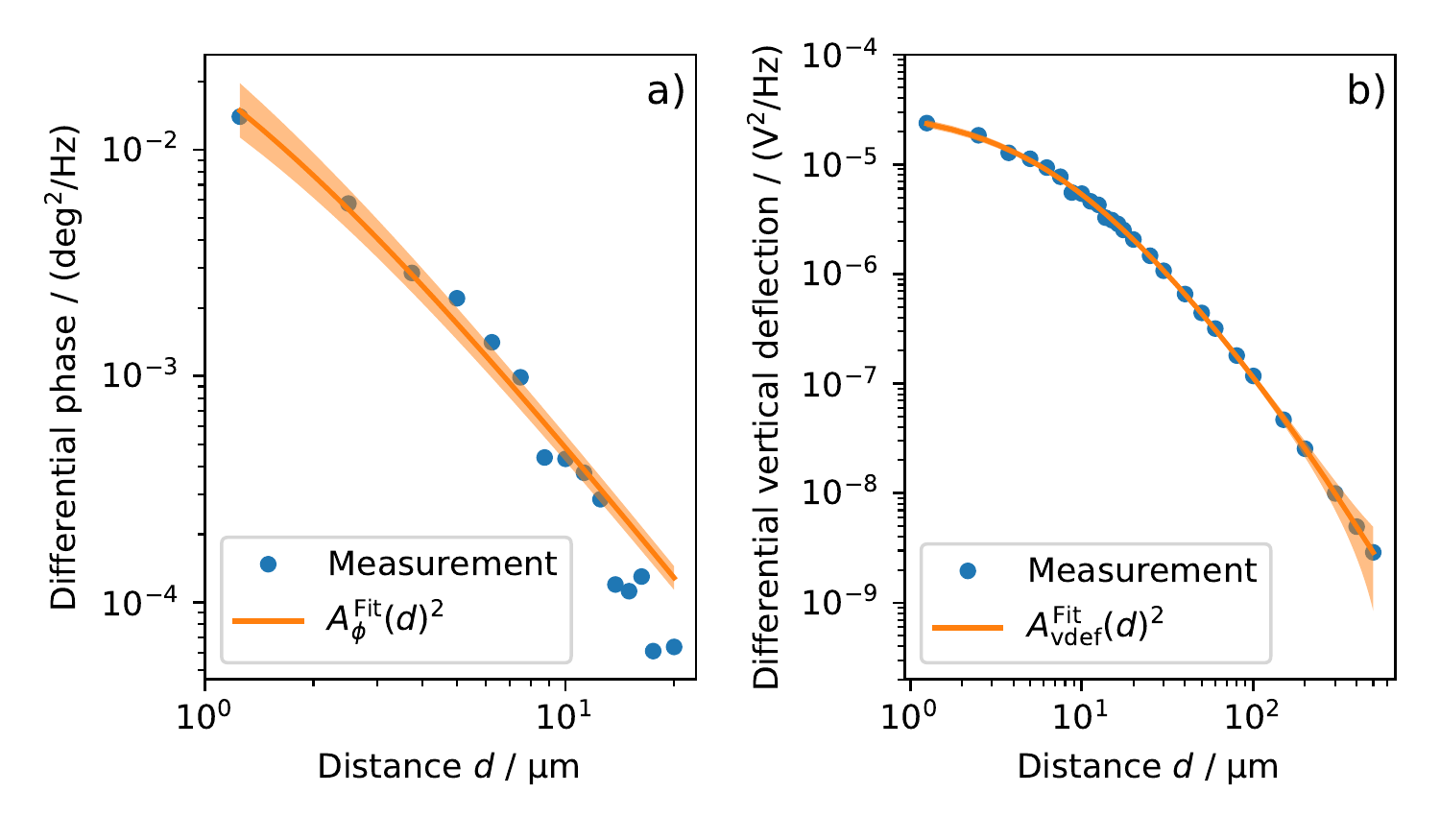}%
 \caption{\textit{Differential phase and the vertical deflection dependence on tip to surface distance.} The distance-dependent amplitudes are shown for the differential phase a) and the vertical deflection (representing the oscillation amplitude) b), recorded by applying a voltage of \SI{10}{\volt} peak to peak at a frequency of \SI{120}{\hertz}. Straight curves represent fits to the data, using the model explained in the text and the supplemental material, the shaded area indicate for one standard deviation. Fit parameters are given by $a_{\phi} = \SI[separate-uncertainty=true]{0.234(12)}{\degree\micro\meter}$, $d_{\phi} = \SI[separate-uncertainty=true]{-0.67(16)}{\micro\meter}$ and $a_{\mathrm{vdef}} = \SI[separate-uncertainty=true]{2.02(95)e-9}{\volt\squared}$, $b_{\mathrm{vdef}} = \SI[separate-uncertainty=true]{2.62(46)e-6}{\volt\squared\micro\meter}$, $c_{\mathrm{vdef}} = \SI[separate-uncertainty=true]{39.36(56)e-3}{\volt\micro\meter}$, $d_{\mathrm{vdef}} = \SI[separate-uncertainty=true]{-6.82(19)}{\micro\meter}$.}
 \label{fig:approach}
\end{figure}

By applying the voltage, the mechanically driven tip experiences a force gradient along its direction of oscillation, which shifts the phase relative to the driving force and changes the oscillation amplitude at the same time. This changes in phase and oscillation amplitude depend on the distance between tip and surface, so that both can be used as a distance probe either to gain topology informations of the surface by recording height informations from a safe distance during scanning or to approach the surface without ever touching it, so that the just mentioned scans can be performed. The approach is performed while applying an AC voltage $V(t) = V_0\sin(\omega_0 t)$. This is necessary to compensate for drifts of the phase and oscillation amplitude while approaching, making it impossible to obtain voltage induced constant changes at distances $\gg \SI{2}{\micro\meter}$. An analytical formula of the expected differential phase and amplitude change $A_\phi(h_0)$, $A_\mathrm{vdef}(h_0)$, as well as a derivation of formulas $A_\phi^\mathrm{fit}(d)$ and $A_\mathrm{vdef}^\mathrm{fit}(d)$, which can be fitted to the acquired data, can be found in the supplementary material (Eq. S6 and S10). The latter are given by:
\begin{align}
A_{\phi}^\mathrm{fit}(d)^2 &= \left(\frac{a_{\phi}}{d-d_{\phi}}\right)^2, \\
A_\mathrm{vdef}^\mathrm{fit}(d)^2 &= a_\mathrm{vdef} - \frac{b_\mathrm{vdef}}{d-d_\mathrm{vdef}} + \left(\frac{c_\mathrm{vdef}}{d-d_\mathrm{vdef}}\right)^2,
\end{align}
with variable fit parameters $a_{\phi}$, $d_{\phi}$, $a_\mathrm{vdef}$, $b_\mathrm{vdef}$, $c_\mathrm{vdef}$ and $d_\mathrm{vdef}$. The tip to surface distance $d$ can be measured directly by the AFM, while the equilibrium distance between tip and conductive surface $h_0$ is not directly accessible in this experiment. One should note that the square of the amplitudes must be fitted, since the power spectral density has been recorded. Those curves fit sufficiently to the experimental data, as can be seen in \autoref{fig:approach} a) and b), represented as solid curves. A mismatch between analytical formula and measured data is obtained for the differential phase at distances above \SI{10}{\micro\meter} due to the increased influence of the force between cantilever and surface at higher distances. Our model only takes into account the force between tip cone and surface, which dominates for distances below \SI{10}{\micro\meter}. In this region, the model matches the data so that it can be used as a calibration curve.

The fitted curves - once derived - then serve as quantitative calibration curves for the approach process. Further measurement runs show only minor deviations, which indicates for a reliable and robust technique. Even a change of the tip shape i.e. due to the picking process, has no major effect on the tip approach procedure. The vertical deflection signal $A_\mathrm{vdef}(h_0)^2$ is used to approach the tip starting at a distance of $\SI{500}{\micro\meter}$. At this large distance, no differential phase can be detected, because the signal scales with $h_0^{-2}$ (see Eq. S5), whereas one term of the vertical deflection scales with $h_0^{-1}$ (see Eq. S9). At $\approx\SI{20}{\micro\meter}$ the phase shift can be detected. Then it is preferably to use $A_\phi(h_0)^2$ to approach the tip close and precise to the surface, as the signal change is higher than the vertical deflection signal.

\begin{figure}[ht]
 \includegraphics[width=\columnwidth]{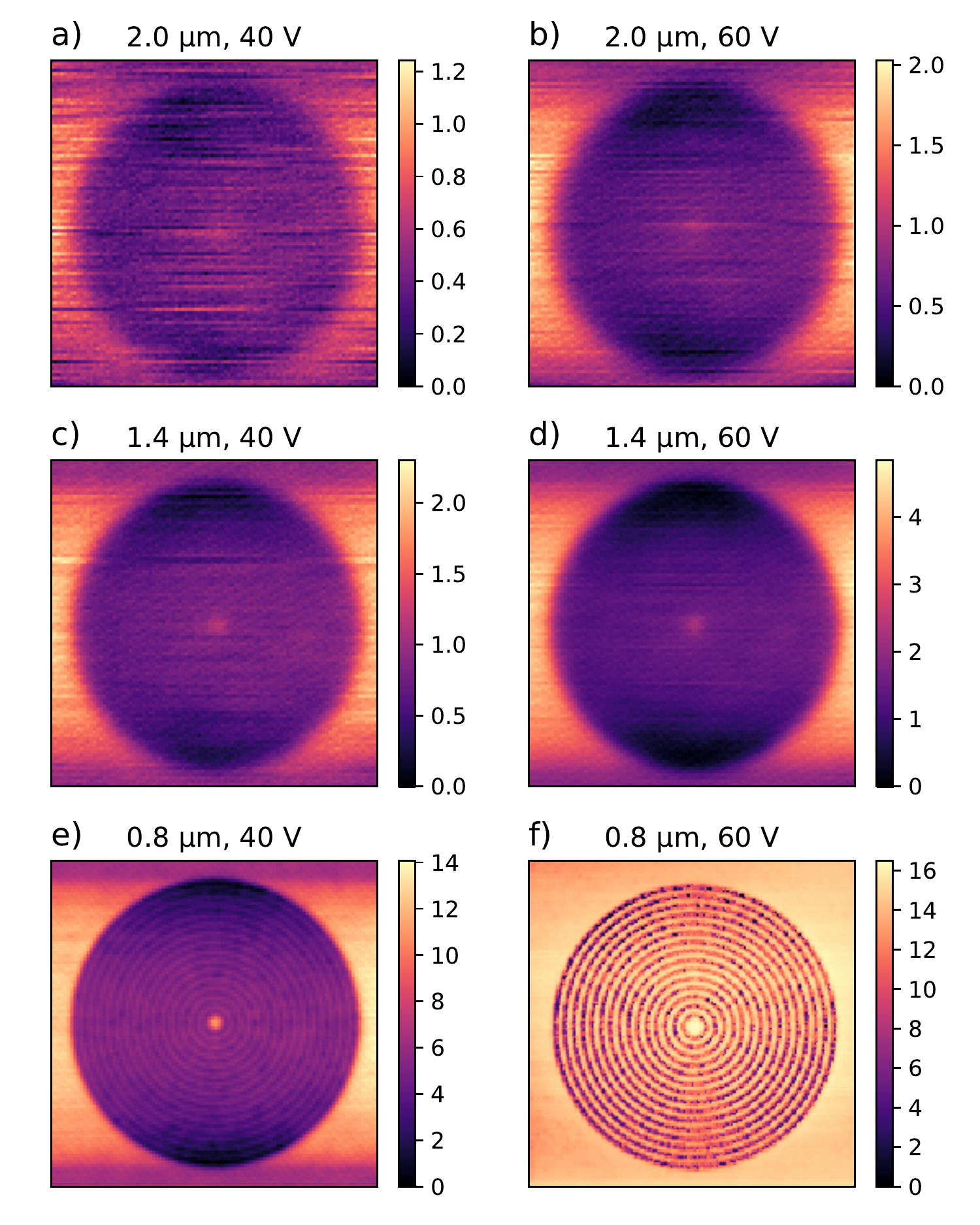}%
 \caption{\textit{Electrostatic force images at different DC voltages and distances.} Color code denotes the measured differential phase shift in $^\circ$. The distance between tip and surface $d$ decreases from top to bottom, the left column was taken with an applied DC voltage of \SI{40}{\volt}, the right one with \SI{60}{\volt}. A reduction of the distance increases the resolution and an increase of the applied voltage increases the signal-to-noise ratio. The scan area is $20\times 20\,\si{\micro\meter\squared}$.}
 \label{fig:EFM}
\end{figure}

Once the non contact approach is done, a DC voltage is applied between tip and sample, which, as described above, causes a constant differential phase due to the occurrence of a force gradient. The distance dependent differential phase recorded during an xy-scan of the AFM tip at a fixed z-position maps the height of the metal surface. \Autoref{fig:EFM} shows these electrostatic force microscope images recorded at different voltages. Increasing the DC voltage at a constant distance increases the signal to noise ratio, as a higher phase change is detected. Lowering the tip to surface distance results in an increased xy-resolution, down to a sub-micron level, which determines the placement accuracy.

\begin{figure}[h]
 \includegraphics[width=\columnwidth]{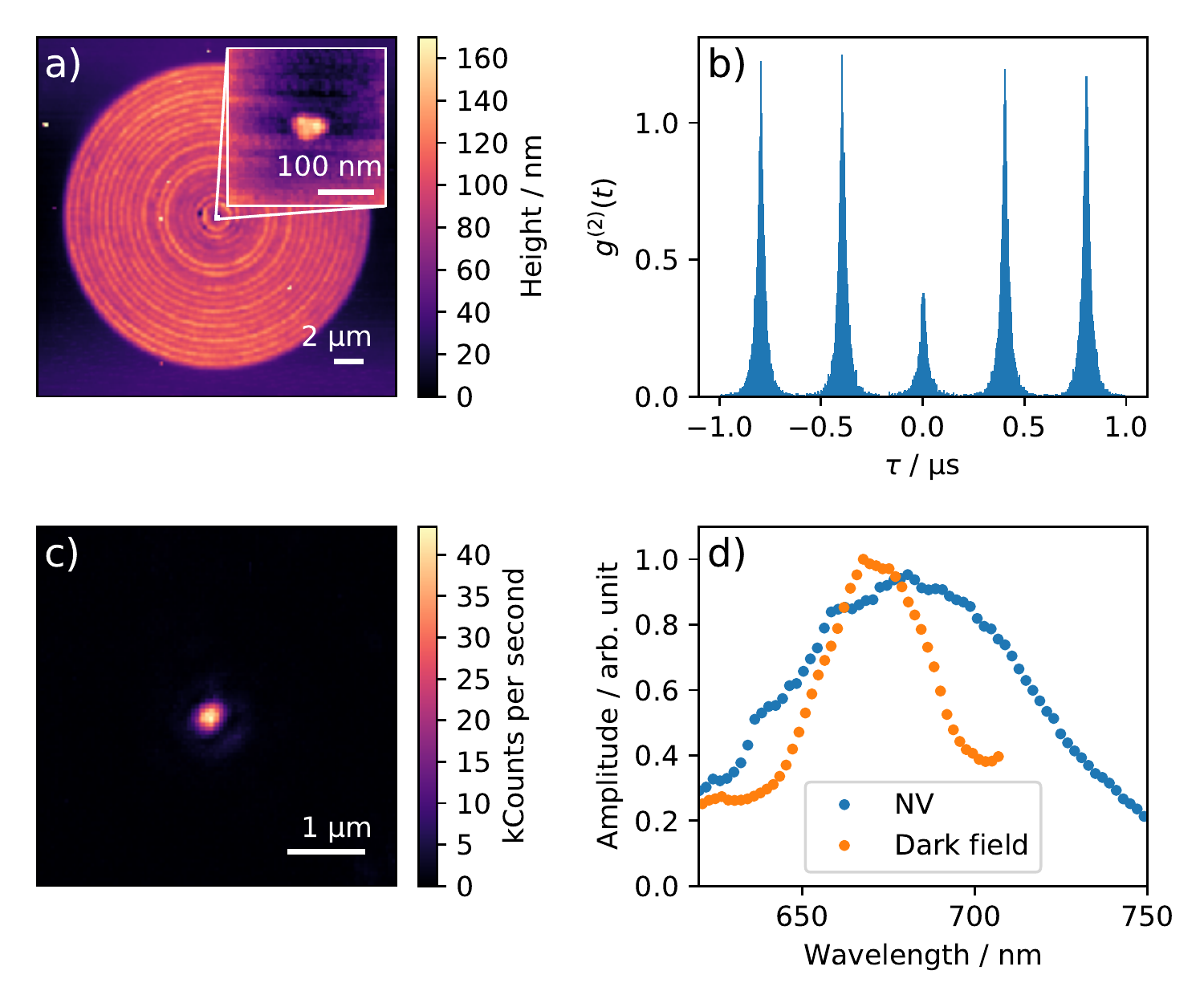}%
 \caption{\textit{AFM, confocal scan and optical characterization of the placed NV.} a) shows AFM scans of the placed nano diamond in the center of the antenna. b) A photon autocorrelation ($g^{(2)}$-function) recorded with a repetition rate of \SI{2.5}{\mega\hertz} shows an antibunching of less than 0.5. c) Confocal scan of the antenna with the nano diamond in the middle. A pulsed laser with a repetition rate of \SI{5}{\mega\hertz} was used and the counts during the first nanosecond after each laser pulse were omitted. In d), the NV spectrum (blue) and a dark field scattering spectrum (orange) are shown.}
 \label{fig:placing}
\end{figure}

After switching off the voltage, the nano diamond is deposited by finally approaching the tip to the surface and applying a force similar to Ref. \cite{Schell2011}. Subsequent nano manipulation reduces the deposition accuracy to the AFM resolution, which in this case correlates with the convolution of tip and diamond size. In this experiment we estimate the typical positioning accuracy to be below \SI{5}{\nano\meter}. In \autoref{fig:placing} a) an AFM-scan shows the final result, i.e. the successfully placed nano diamond in the center of a bullseye antenna. An optical confocal measurement of the same antenna reveals only a single diffraction limited spot (see \autoref{fig:placing} c). With a NA 0.9 objective lens, $300\,\mathrm{kCounts/s}$ in saturation were detected. The excitation laser (Solea, PicoQuant) operated with a repetition rate of \SI{40}{\mega\hertz}, at a central wavelength of \SI{540}{\nano\meter}, a wavelength span of \SI{15}{\nano\meter} and a CW power of \SI{1.1}{\milli\watt}. A $g^{(2)}$-function taken from this fluorescence signal shown in \autoref{fig:placing} b) confirms the emission of non-classical light stemming predominately from a single emitter \cite{Schroder2011}. This result nicely shows the operation of the antenna, i.e. efficient collection of photons, once the emitter is precisely positioned. A drawback of our chosen emitter, the NV center, is its broad emission spectrum extending from \SI{640}{\nano\meter} to $\SI{750}{\nano\meter}$ (see fluorescence spectrum in \autoref{fig:placing} d). However, the bullseye antenna accounts for this by its broad optical response. In order to show this, we measured the dark field antenna spectrum, by collimating a light beam onto the antenna region. The direct reflection was filtered out and residual light stemming from the center was guided into the spectrometer. In this way, the inverse antenna operation was simulated. The resulting spectrum showed in d) fully overlaps with the NV fluorescence spectrum acquired from the positioned particle. Furthermore, the scattering efficiency profile indicates for potentially best collimation at \SI{670}{\nano\meter}. The measurements represent one example how our method of functionalization can be applied successfully to assemble an optimal structure for photon collection from NV centers in nano diamonds.

In summary, we presented a fast and versatile method that allows the deterministic placement of individual nano particles on opaque conductive substrates. After an initial tip-sample calibration, our method provides an effective and reproducible way to functionalize any kind of plasmonic or photonic structures realized on conductive surfaces, with a sub-micrometer precision limited only by the employed AFM.
\\
\\
This work was supported by the Einstein Foundation Berlin project "ActiPlAnt" and the German Ministry of Education and Research (BMBF) project "NANO-FILM".

\bibliography{references}

\end{document}


\title{
Supplementary material: Fast and accurate functionalization of opaque conductive samples with single nano particles
}
\author{Niko Nikolay$^{1,2}$}
\author{Nikola Sadzak$^{1,2}$}
\author{Alexander Dohms$^{1,2}$}
\author{Boaz Lubotzky$^3$}
\author{Hamza Abudayyeh$^3$}
\author{Ronen Rapaport$^3$}
\author{Oliver Benson$^{1,2}$}

\affiliation{
 $^1$ AG Nanooptik, Humboldt-Universität zu Berlin,
 Newtonstraße 15, D-12489 Berlin, Germany
}
\affiliation{
 $^2$ IRIS Adlershof, Humboldt-Universität zu Berlin,
 Zum Großen Windkanal 6, 12489 Berlin, Germany
}
\affiliation{
 $^3$ The Racah Institute of Physics, The Hebrew University of Jerusalem,
 Jerusalem 9190401, Israel
}

\date{\today}

\maketitle

\section{An oscillating cantilever in a field gradient}
\label{sec:second}

When an AFM tip above a conductive surface is mechanically driven with a frequency $\omega_\mathrm{mechanic}$ close to the cantilever resonance frequency $\omega_\mathrm{C}$, it encounters a force gradient along its oscillation direction. This force gradient, depending on the applied voltage and the tip-sample distance, induces a change $\mathit{\Delta}\phi$ in the oscillation phase and a change $\mathit{\Delta} A_\mathrm{osc}$ in the oscillation amplitude.

\begin{figure}[ht]
 \includegraphics[width=\columnwidth]{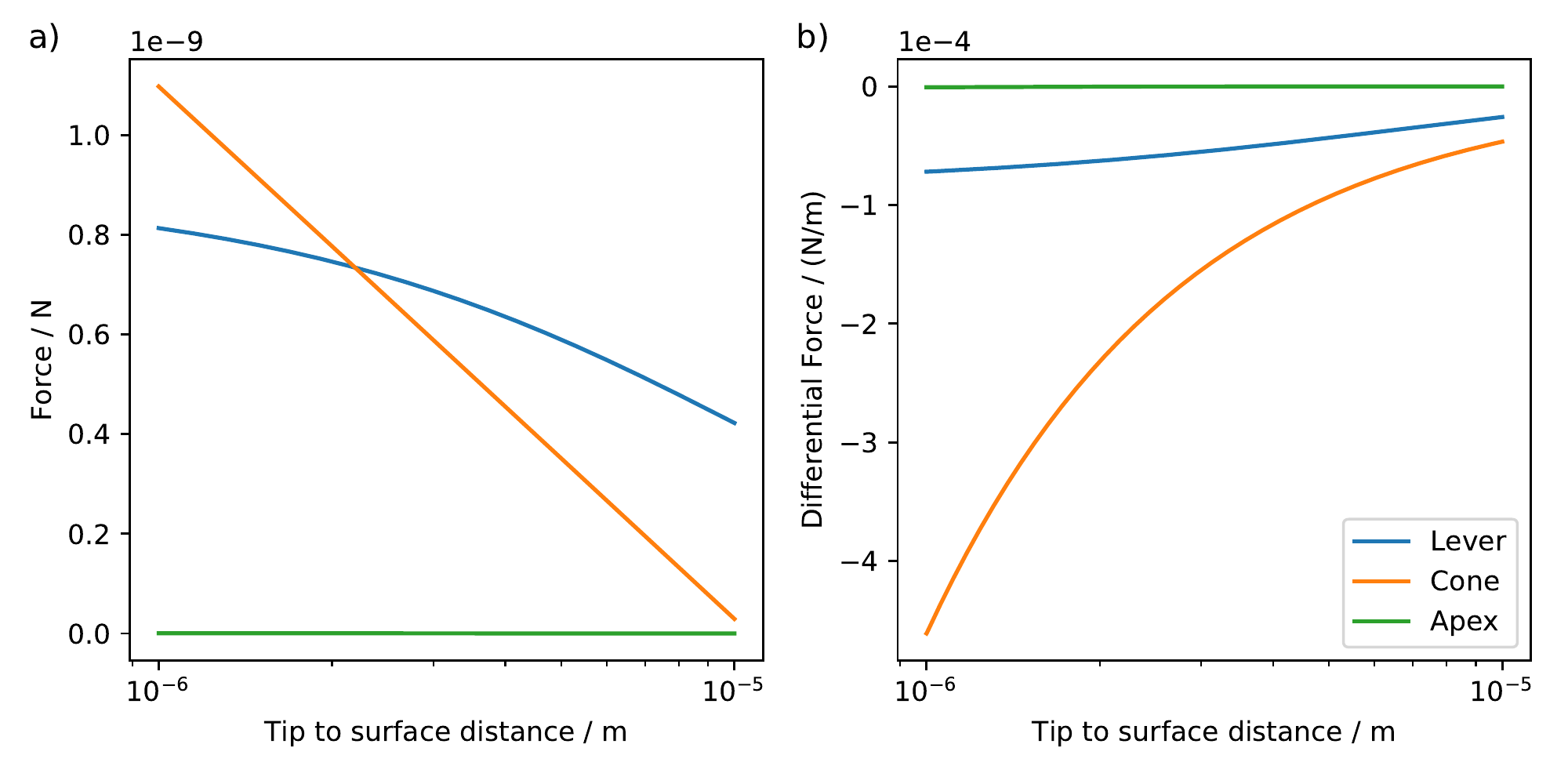}
 \caption{\textit{Electrostatic forces between tip and surface.} Force a) and differential force b) at different tip-to-surface distances are shown, calculated with the analytical formulas taken from Ref. \cite{Colchero2001} with a constant Voltage of \SI{5}{\volt}, and geometric parameters for the HQ:NSC35/Pt (MikroMasch), i.e. lever width of \SI{35}{\micro\meter}, lever length of \SI{130}{\micro\meter}, lever tilt angle of \SI{10}{\degree}, cone length of \SI{15}{\micro\meter}, tip end height of \SI{60}{\nano\meter}, cone opening angle of \SI{40}{\degree} and a tip radius of \SI{30}{\nano\meter}.}
 \label{fig:force}
\end{figure}

A general, analytical equation for the force experienced by an AFM tip $F_\mathrm{V}$ subject to an electric potential $V$ is given in Ref. \cite{Colchero2001}. Here, the total force is described as the sum of three contributions, namely the force experienced by the cantilever $F_\mathrm{lever}$, the tip cone $F_\mathrm{cone}$ and the tip apex $F_\mathrm{apex}$, so that $F_\mathrm{V}(h) = F_\mathrm{lever}(h) + F_\mathrm{cone}(h) + F_\mathrm{apex}(h)$. \autoref{fig:force} shows the respective force components $F_i(h)$ (a) and its differentials $\partial F_i(h) / \partial h$ (b) as a function of the tip to surface distance $h$.

As can be seen in \autoref{fig:force} b), the most prominent force gradient in the near-surface approaching range (that is \SIrange{1}{10}{\micro\meter}) is caused by the cone. An approximation of this force (taken from Ref. \cite{Colchero2001}), under the assumption that the tip end height $\delta \ll h$, is given by
\begin{equation}
\label{eq:fcone}
F_\mathrm{cone} \approx \frac{4\pi\epsilon V^2}{(\pi-\Theta_\mathrm{tip})^2} \ln \left( \frac{h_\mathrm{cone}}{h} \right) = F_0 \ln \left( \frac{h_\mathrm{cone}}{h}\right),
\end{equation}
with the potential $V$, the opening angle $\Theta_\mathrm{tip}$ and the cone height $h_\mathrm{cone}$.

\begin{figure}[ht]
 \includegraphics[width=\columnwidth]{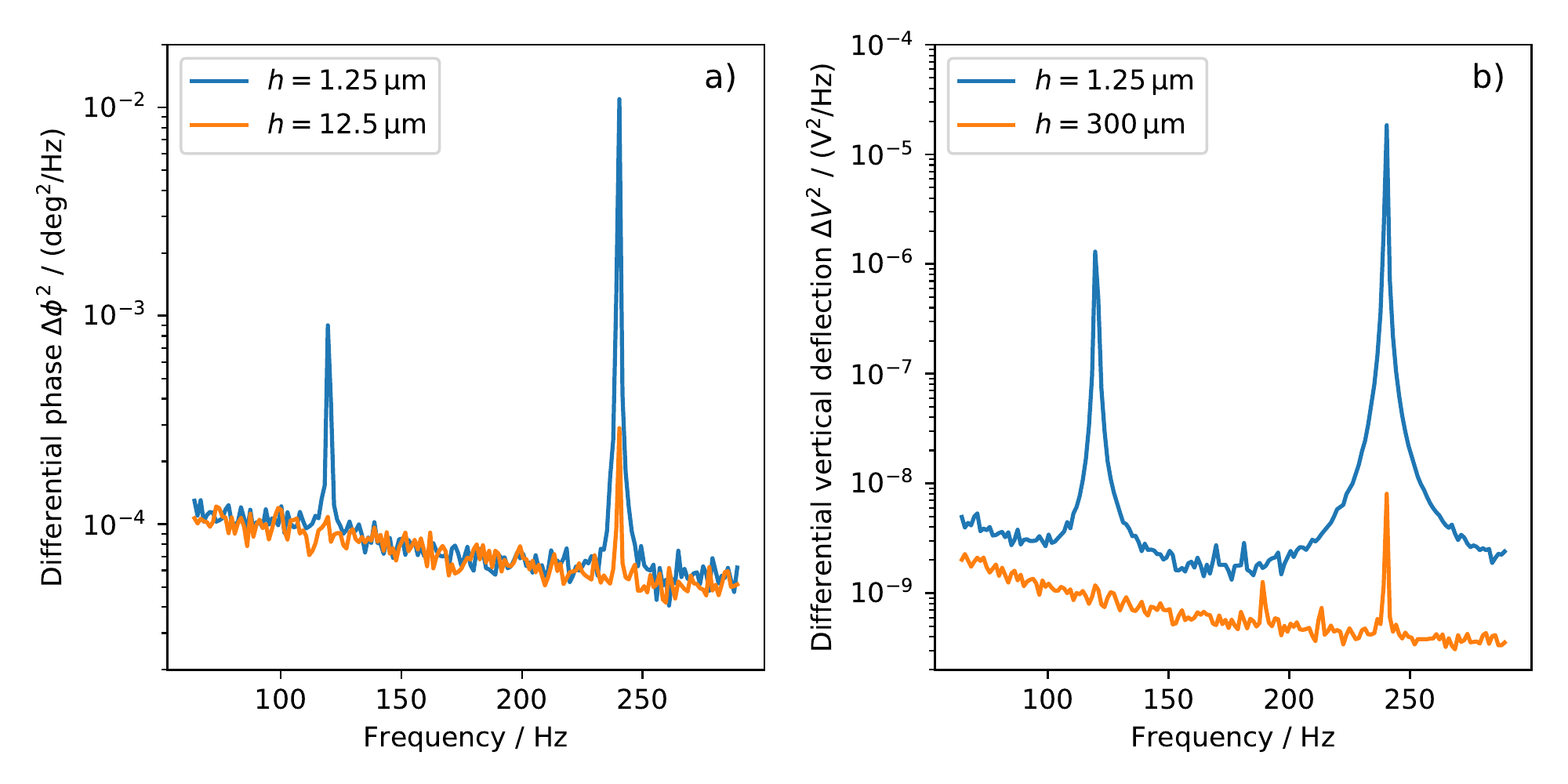}
 \caption{\textit{Differential phase and vertical deflection spectra.} Differential phase a) and vertical deflection spectra b) at different tip-to-surface distances are shown. A driving voltage of $2V_0=\SI{10}{\volt}$ peak to peak at $\omega_0=\SI{120}{\hertz}$ was applied. The fundamental and the second harmonic oscillations of the cantilever are observed in both of the spectra, as predicted by \autoref{eq:phasediff_fourier}.}
 \label{fig:approach}
\end{figure}

\Autoref{fig:approach} a) and b) show the power spectral density of the measured differential phase $\mathit{\Delta}\phi^2$ and differential vertical deflection signals $\mathit{\Delta} V^2$ when applying an alternating voltage. The differential vertical deflection signal is the differential voltage detected by the AFM four-quadrant photo diode and transforms linearly to the cantilever oscillation amplitude $\mathit{\Delta} A_\mathrm{osc}$. Both values were measured with the JPK NanoWizard setup and the built in software. Resonances at the fundamental \SI{120}{\hertz} and the second harmonic \SI{240}{\hertz}, can be clearly identified. The shown frequency and height dependence is relevant for the functionalization experiment and will therefore be discussed in the next two chapters.

\section{Derivation of an analytical formula for the phase-dependency}
\label{sec:phase}

A formula for calculating the phase change caused by a gradient field, which applies to this situation, is given in Ref. \cite{Lilliu2013}:
\begin{equation}
\mathit{\Delta}\phi(h_0, t) = - \frac{Q}{k}\frac{\partial F(h)}{\partial h} = - \frac{1}{2}\frac{Q}{k}\left[\frac{\partial^2C(h)}{\partial h^2}\left(V(t)-V_\text{CPD}\right)^2\right]_{h=h_0},
\label{eq:main}
\end{equation}
with the cantilever quality factor $Q$, the cantilever elastic constant $k$, the distance between tip and conducting surface (or height) $h$, the equilibrium distance $h_0$, the force $F(h)$, the externally applied voltage $V(t)$, the contact potential difference $V_\text{CPD}$ and the tip to surface capacity $C(h)$. An alternating voltage $V(t) = V_0\sin(\omega_0 t)$ induces phase variations at specific frequencies, which can be determined by the Fourier transform of \autoref{eq:main} calculated under the condition $\omega_0 \ll \omega_\mathrm{C}$:
\begin{equation}
\label{eq:phasediff_fourier}
\mathit{\Delta}\hat{\phi}(h_0,\omega)\Bigr|_{0<\omega<\omega_\mathrm{C}} = \sqrt{\frac{\pi}{2}}\frac{Q}{k}\frac{\partial^2C}{\partial h^2}\Bigr|_{h=h_0} \left(\iu V_0 V_\text{CPD}\delta(\omega-\omega_0)+\frac{1}{4}V_0^2\delta(\omega-2\omega_0)\right).
\end{equation}
According to this expression, spectral components of the phase variations are expected at the frequency $\omega_0$ (fundamental) and $2\omega_0$ (second harmonic), with the amplitudes being dependent on the distance $h_0$. We have neglected the influence of the applied voltage bending the cantilever and the influence of the much faster tip oscillation due to the mechanical driving, both of which would affect the capacity. We justify our approximation by noting that in our system, the oscillation amplitude due to the mechanical driving is below \SI{50}{\nano\meter}, and the bending due to the applied voltage of \SI{10}{\volt} peak to peak with a frequency of $\nu = \SI{120}{\hertz}$ is below \SI{5}{\nano\meter} even at a distance $h_0$ of \SI{1.25}{\micro\meter}. Both values are small compared to $h_0$, and hence have no influence on the capacity.

In order to safely approach the tip within few micrometers from the surface, we keep track of the second harmonic component amplitude to probe the distance, since it depends only on the external voltage and the tip-sample separation, while material specific parameters that could change locally do not contribute to the signal. The distance dependent amplitude of the second harmonic extracted from \autoref{eq:phasediff_fourier} is given by
\begin{equation}
\label{eq:aphi}
A_\phi(h_0) = \frac{1}{4}\sqrt{\frac{\pi}{2}}\frac{Q}{k}\frac{d^2C}{dh^2}\Bigr|_{h=h_0}V_0^2 = \frac{1}{2} \sqrt{\frac{\pi}{2}} \frac{Q}{k} \frac{\partial F_\mathrm{V}(h)}{\partial h},
\end{equation}
with the force $F_\mathrm{V}(h) = 1/2\,\partial C(h)/\partial h\, V_0^2$. The force calculation from \autoref{eq:fcone} combined with \autoref{eq:aphi} gives the distance dependence of the differential phase
\begin{equation}
A_\phi(h_0) = \sqrt{\frac{\pi}{8}} \frac{Q}{k} F_0 \frac{1}{h_0}.
\end{equation}
This formula can be used to calculate the tip to surface distance from non-contact differential phase measurements. We introduce two fit parameters $a_{\phi}$, $d_{\phi}$ to account for our specific experimental condition (i.e. the coating of the substrate with PMMA, Polymethylmethacrylate) and for a remaining effect of $F_\mathrm{lever}$:
\begin{equation}
\left[A_{\phi}^\mathrm{fit}(d)\right]^2 = \left(\frac{a_{\phi}}{d-d_{\phi}}\right)^2,
\end{equation}
where $d$ is the distance between tip and sample surface that can be measured by the AFM, while the distance between tip and the conductive surface beneath (height) is not directly accessible in this experiment. One should note that in our case we fit the square of the amplitudes, since we record power spectral density data.

\section{Derivation of an analytical formula for the vertical deflection}
\label{sec:vdef}

We now come back to the distance dependence of the vertical deflection shown in \autoref{fig:approach} b). In the supporting material of Ref. \cite{Lilliu2013}, the response of a mechanically driven cantilever in a force field is analytically calculated. The inverse of the derived transfer function $\widetilde{W}$ gives the oscillation amplitude $A_\mathrm{vdef}$. As in the previous section, the square of the amplitude is of particular interest:
\begin{equation}
A_\mathrm{vdef}^2(h_0) = \frac{1}{\widetilde{W}^2} = m^2\omega^4 + \left( b^2-2\left(k-\frac{\partial F(h)}{\partial h}\Bigr|_{h=h_0} \right) m\right)\omega^2+\left( k - \frac{\partial F(h)}{\partial h}\Bigr|_{h=h_0} \right)^2,
\end{equation}
with the effective mass $m$, frequency $\omega$, damping constant $b$, stiffness constant $k$ and the distance dependent force $F(h)$. By substituting $A = m^2\omega^4+b^2\omega^2-2mk\omega^2+k^2$ and $B=2M\omega^2-2k$, this formula becomes:
\begin{equation}
\label{eq:Isimple}
A_\mathrm{vdef}(h_0)^2 = A+B\frac{\partial F(h)}{\partial h}\Bigr|_{h=h_0} + \left(\frac{\partial F(h)}{\partial h}\Bigr|_{h=h_0} \right)^2.
\end{equation}
In our measurement range as from \autoref{fig:force} b), the force acting on the AFM cantilever is mainly due to the tip cone and surface interaction, as discussed in the previous section. Since the frequency of the applied voltage $\omega$ is much lower than the cantilever resonance frequency $\omega_\mathrm{C}$, the cantilever will just follow the applied electric force without any resonance effect. We measure the oscillation amplitude, which is therefore induced by the applied voltage $V_0$. Inserting the analytic expression for the force given by \autoref{eq:fcone} into \autoref{eq:Isimple} gives:
\begin{equation}
A_\mathrm{vdef}(h_0)^2 = A-B F_0 \frac{1}{h_0} + F_0^2 \frac{1}{h_0^2}.
\end{equation}
For the same reasons as in the previous section, we introduce fit parameters $a_\mathrm{vdef}$, $b_\mathrm{vdef}$, $c_\mathrm{vdef}$, $d_\mathrm{vdef}$ and change to the tip to surface distance $d$:
\begin{equation}
A_\mathrm{vdef}^\mathrm{fit}(d)^2 = a_\mathrm{vdef} - \frac{b_\mathrm{vdef}}{d-d_\mathrm{vdef}} + \left(\frac{c_\mathrm{vdef}}{d-d_\mathrm{vdef}}\right)^2.
\end{equation}

\bibliography{references}